\documentclass[10pt]{iopart} 

\newcommand{\be}{\begin{eqnarray}}
\newcommand{\ee}{\end{eqnarray}}
\newcommand{\raf}[1]{(\ref{#1})}

\usepackage{epsf}  
\begin{document}

\article{Strange Quark Matter 2003}{Strangeness at SIS energies}

\author{Volker Koch
}

\address{Nuclear Science Division, Lawrence Berkeley National Laboratory,
  Berkeley, CA 94720, USA}

\begin{abstract}
In this contribution we discuss the physics of strange hadrons in low energy 
($\simeq 1-2 \, \rm AGeV$) heavy ion collision. In this energy range the
relevant strange particle are the kaons and anti-kaons. The most interesting
aspect concerning these particles are so called in-medium modifications. We
will attempt to review the current status of understanding of these in medium
modifications. In addition we will briefly discuss other issues related with
kaon production, such as the nuclear equation of state and chemical
equilibrium.
\end{abstract}


\section{Introduction}
The production of strange particles in heavy ion collisions has been studied
extensively for all available energies. While fits to particle yields
\cite{redlich} seem to indicate a common systematic for particle production
over 
the entire energy range, as far as strange particle production and
dynamics is concerned there is, at least from the theoretical point of view, 
a conceptual difference between the low energies and the ultra-relativistic
regime. At low energies, the system is dominated by (non-strange) baryons,
whereas for SPS-energies ($\sqrt{s} = 20 \,\rm GeV$) 
and higher the system is temperature 
dominated. In the limit of perfect flavor SU(3) symmetry a system at
zero baryon chemical potential and finite temperature preserves 
the SU(3) flavor symmetry. Contrary to that, a system at low
temperature, but large chemical potential,
breaks the SU(3) flavor symmetry {\em explicitely}. The extreme example is a
cold nucleus which consists only of protons and neutrons, but does not
contain any hyperons. As a consequence, slow kaons and anti-kaons
feel, contrary to pions,
repulsive and attractive mean field potentials, respectively. This is analogous
to pions in neutron matter, 
where the in-medium potentials for positive and
negative
pions are different due to the explicit {\em isospin} breaking of neutron
matter. As the bombarding energy is increased, and the system becomes more
temperature dominated, these effects are less relevant and more difficult to
extract experimentally.
Thus, low energy heavy ion collision provide a unique environment to study 
the SU(3) flavor dynamics and its effects on the strangeness sector at 
densities above the nuclear matter ground state density. 

Historically, the interest in  kaon production at low energies 
has been motivated by the
search for the nuclear equation of state \cite{aichelin_ko}. 
Since at these energies kaons are produced
either sub or near threshold, kaon production should be sensitive to the
equation of state. This approach, however, can only work if all
other forces on the kaons, such as mean fields, are well understood. By now 
a rather consistent picture of kaon production has emerged, and the
resulting equation of state is in agreement with the analysis of flow
observables. 

In recent years the production of anti-kaons has received considerable 
attention. Especially the connection to the structure of compact (neutron) 
stars, with the possible formation of a kaon ($K^-$) condensate
\cite{kaplan,brown}, generated interest in anti-kaon production at high
densities. Unfortunately, the environment generated in a heavy ion collision
is different from that of a neutron star. The system is a finite temperature
($T \simeq 70 \, \rm MeV$) and the kaons observed in experiment have finite
momenta. Therefore, a careful
extrapolation to the conditions inside a compact star -- zero temperature and
momentum -- needs to be carried out. This, as we shall discuss, introduces
quite some ambiguities and model dependencies which require further detailed
investigation.

This contribution is organized as follows. First, we will discuss the
conceptual differences between kaon and anti-kaon production.
Then we will briefly summarize the present understanding of
kaon ($K^+$) production, which is rather non-controversial. A large part of
this contribution will be devoted to anti-kaon production, because here the
interpretation of the data is not yet settled. Also, the in-medium properties
for anti-kaons present a nice and possibly tractable example for in-medium
modifications of hadronic properties. Before we conclude, we will briefly
mention how kaon production can also be utilized to address the question of
equilibration in these collisions.

Throughout this contribution we will concentrate on the SIS energy
regime. While some of the effects are also seen at the lower AGS energies
($\simeq 4 \, \rm AMeV$), there, temperature effects wash out most of the
interesting dynamics responsible for  the anti-kaon in-medium properties.

\section{Kaon and anti-kaon production in a density dominated environment}
Initially, the properties of kaons in dense matter 
has been discussed in the framework of chiral perturbation theory. In
ref. \cite{kaplan} the leading order interaction between a kaon and nucleon
has been given by
\be
\delta {\cal L} = - \frac{3}{8 f^2} 
(\bar{K} i \stackrel{\leftrightarrow}{\partial} K) 
\bar{N} \gamma_0 N
+ c (\bar{K} K)(\bar{N}{N})
\ee
The first term reflects the vector interaction analogous to the
Weinberg-Tomozawa term know from pion physics. This term is repulsive for
kaons and attractive for anti-kaons. The second term, which is due to explicit
symmetry breaking, is attractive for both, kaon and anti-kaon. The coefficient
of this term is much less known than that of the vector interaction. The
resulting interaction is  repulsive for the kaons, in agreement with
experiment, but attractive for anti-kaon, which is not supported by
$K^-$-nucleon scattering data which also give a repulsive scattering length
\cite{martin}. The reason for this discrepancy is the existence of the
$\Lambda(1405)$ resonance right below the $\bar{K} -N$-threshold,
which  couples strongly to the the $\bar{K} -N$ system. Thus, re-summed chiral
perturbation theory is required in order to describe the  $\bar{K} -N$ system
correctly\cite{weise,oset}. 
As a result, the $\Lambda(1405)$ is generated as a bound state,
which in fact has already been proposed almost forty years ago \cite{dalitz}.

The importance of the $\Lambda(1405)$ illustrates the essential difference
between the anti-kaon--nucleon and kaon--nucleon system. While the former is
dominated by resonances, the latter does not show any significant structure
in the scattering amplitude, since the strange anti-quark cannot be
absorbed into a baryon resonance. This difference will also affect the in
medium properties of kaons and anti-kaons. Since the kaon-nucleon interaction
is non-resonant, we expect that the impulse approximation should describe the
kaon-nucleus system as well as kaon propagation in heavy ion collision rather
well. We thus expect that a straightforward mean field description should be
reasonable for kaons in nuclear matter. This is different for the
anti-kaon. Here we expect strong mixing with resonance-hole states, similar to
pion propagation in nuclear-matter. Thus a detailed coupled channel treatment
is called for. 

\subsection{Kaon properties in dense matter}
Let us first discuss the simpler case of kaons in matter. As already pointed
out, we expect that kaons in matter should be reasonably well described by a
mean-field obtained via the impulse approximation. And indeed this works rather
well for the description of kaon-nucleus scattering
experiments \cite{ernst}. Since the scattering amplitude is non-resonant,
corrections due to finite temperature should be small, and thus, the impulse
approximation should provide a reasonable estimate for the kaon mean field
potential. Using the measured scattering length one obtains
\be
U_{\rm opt} \simeq 20 \, {\rm MeV} \frac{\rho}{\rho_0}
\label{eq:k_pot}
\ee
which is slightly repulsive. 

\begin{figure}[htb]
\epsfxsize=0.8 \textwidth
\centerline{\epsfbox{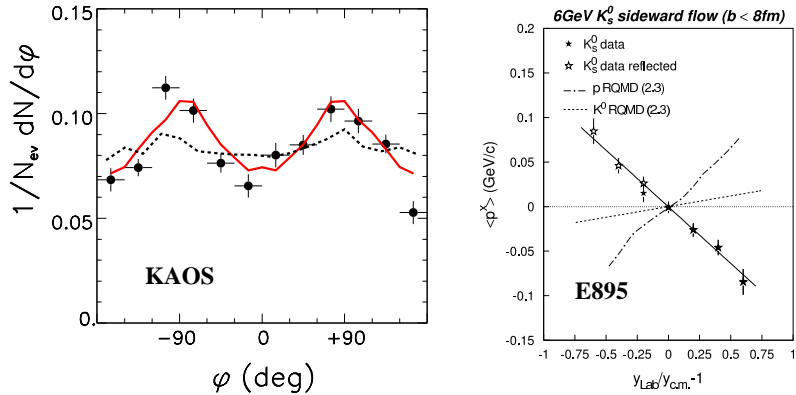}}
\caption{Left panel: Kaon squeeze out obtained by the KAOS collaboration
  \cite{kaos_flow}. The lines are transport calculations \protect\cite{li_buu} 
with (full) and without(dashed) a kaon potential. Right panel: 
Kaon sideways flow at AGS
  energies obtained by the E895 collaboration \protect\cite{e895_flow}. Also
  shown   are the results of transport calculations using the RQMD model.}
\label{fig:k_flow}
\end{figure}

As predicted by Ko et al. \cite{ko_li}, this
repulsion should be visible in kaon flow observables and, as can be seen in
Fig. \ref{fig:k_flow}, the data clearly favor a repulsive mean field for the
kaons. For SIS energies, where transport calculations are more reliable, the
above mean field potential \raf{eq:k_pot} leads to a quantitative agreement
with the data \cite{kaos_flow,fopi_flow} for the so-called squeeze out 
and the sideways flow \cite{ko_li,cassing,wang}. The
repulsive potential for the kaons is even seen at AGS energies (left panel of
Fig. \ref{fig:k_flow}). Strong flow away from the nucleons is observed
\cite{e895_flow}. 
Thus the presence of a repulsive mean field for kaons in
nuclear matter up to twice nuclear matter density seems to be called for.
 
In addition, the resulting kaon yield as a function of system size can be
utilized to extract information about 
the equation of state. Again, the data are
reproduced, and,  provided the kaon mean field is taken into account,  the
resulting equation of state is consistent with other observables such as
nucleon flow \cite{kaos_eos,fuchs}.

\subsection{Anti-kaons in dense matter}

\begin{figure}[htb]
\epsfxsize=0.8 \textwidth
\centerline{\epsfbox{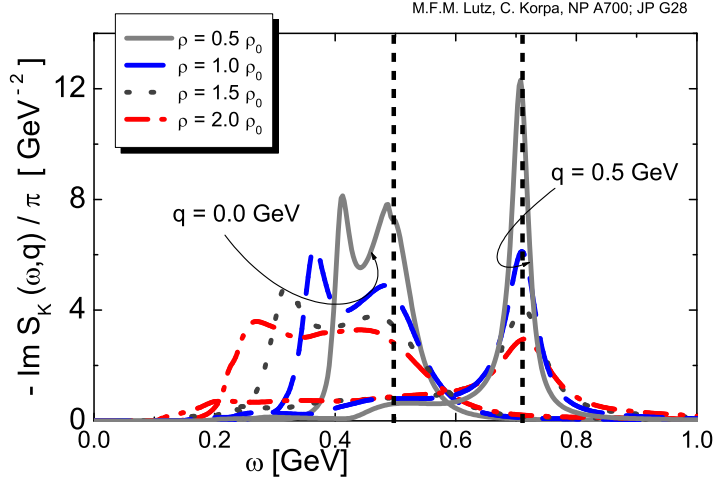}}
\caption{Example for an in-medium anti-kaon spectral function \cite{lutz}. The
  dashed vertical lines indicate the energies for a free anti-kaon at the given
  momentum.}
\label{fig:lutz_spec}
\end{figure}

While the properties of kaons in dense matter seem to be rather well
described by a simple repulsive mean field potential based on the impulse
approximation, the situation for anti-kaons is much more complicated; the main
reason being the aforementioned coupling to baryon resonances. The effect of
this more complex dynamics becomes already apparent in the spectroscopy of
kaonic atoms, where very low nuclear densities are probed. Whereas the
$\bar{K}-N$-amplitude is repulsive, kaonic atoms call for an attractive mean
field \cite{gal}. 
Thus, already at densities well below nuclear matter density the
impulse approximation fails. The observed attraction in kaonic atoms can be
naturally explained if one assumes that the $\Lambda(1405)$ resonance is indeed
a $\bar{K}-N$-bound state \cite{koch_1}. 
In this case, Pauli-blocking of the  $\bar{K}-N$
intrinsic wavefunction shifts this state up in energy above the $\bar{K}-N$
threshold before it then
dissolves. As a result, the  underlying {\em attractive} $\bar{K}-N$ 
interaction, which was responsible for the binding of the $\Lambda(1405)$, 
is revealed. The resulting attractive potential is then in reasonable agreement
with the findings from kaonic atoms, which by themselves are somewhat model
dependent \cite{gal}. This is similar to proton-neutron
scattering in the deuteron channel, where the scattering amplitude is
repulsive due to a bound state just below threshold. 
Again, the true interaction
is attractive and is only revealed in an atomic nucleus where the deuteron is
dissolved due to Pauli-blocking. While this idea has first been developed based
on an effective $\bar{K}-N$ interaction \cite{koch_1}, more refined
calculations employing re-summation techniques using  chiral Lagrangians find
the same result \cite{weise_2,oset_2,lutz_2}.
But the $\Lambda(1405)$ is only one of
many resonances the anti-kaon can couple to, 
and a consistent treatment requires
the coupling of all these channels. As a result a rather complex spectral
distribution for the excitations with the quantum numbers of the anti-kaon
emerges. An example of a $K^-$ spectral function based on a chiral Hamiltonian 
is shown in Fig.\ref{fig:lutz_spec}. In this model not only the
$\Lambda(1405)$ but many other resonances are generated dynamically.
The spectral function for zero momentum kaons 
shows the mixing of the kaon with resonance-hole states 
and, as the density is increased, the strength broadens considerably.
Obviously, a simple quasi-particle plus mean field description of the
anti-kaon will oversimplify this situation considerably. 

\begin{figure}[htb]
\epsfxsize=0.8 \textwidth
\centerline{\epsfbox{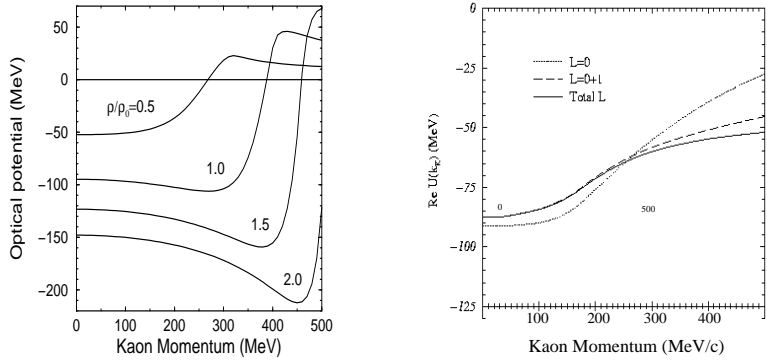}}
\caption{Anti-kaon optical potential. Left from ref. \cite{schaffner}, 
right from  ref. \cite{tolos}}
\label{fig:k_pots}
\end{figure}

In addition, a resonance dominated interaction also should lead to strong
energy/momentum dependencies for the spectral distribution. Furthermore, one
would also expect that the strong density dependent mixing of states should
modify the transition rates between the states. In the simple model of
ref. \cite{koch_1}, where the essential in-medium effect is due to 
Pauli-Blocking
of the $\Lambda(1405)$ one immediately expects that for kaons with 
momenta that 
are large compared to the Fermi momentum of the nuclear matter, the free
(slightly repulsive) $\bar{K}-N$ interaction should re-emerge. This 
indeed is
found as shown in Fig.\ref{fig:k_pots} (left panel). The right panel of
Fig.\ref{fig:k_pots} shows the results of a calculation based on a different,
more sophisticated interaction \cite{tolos}. Again with increasing momentum,
the $\bar{K}$ potential becomes less attractive. But in this case, some net
attraction remains even for momenta of the order of $500 \, \rm MeV$.   

The momentum dependence of the $\bar{K}$ in-medium modifications can also be
read off from the spectral distribution shown in Fig.\ref{fig:lutz_spec}. The
set of curves on the right shows the spectral function for a anti-kaon momentum
of $q= 500 \, \rm MeV$. Not only are the distributions much less broadened but
they all peak at the energy of a free kaon. Thus, the effective kaon potential
is essentially zero, whereas at zero momentum almost all the strength
accumulates 
below the free value. Thus, this calculation also predicts a
reduced attraction  as the momentum of the kaons is increased.  

\begin{figure}[htb]
\epsfxsize=\textwidth
\centerline{\epsfbox{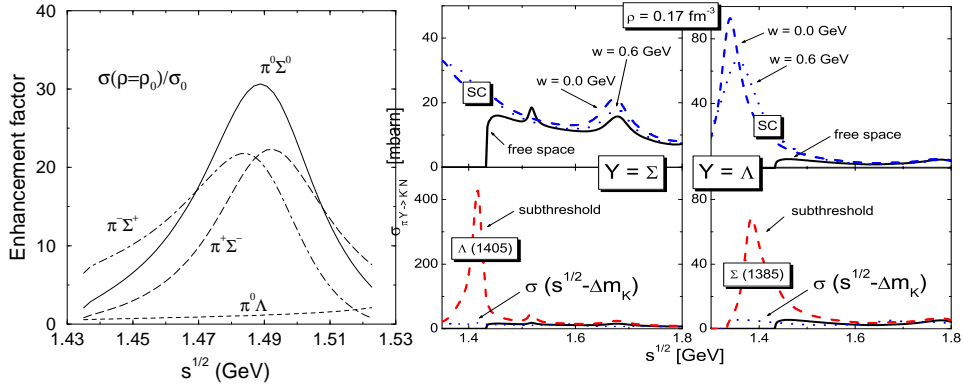}}
\caption{In-medium cross sections. Left from ref. \cite{schaffner}, right from
  ref. \cite{lutz_habil}} 
\label{fig:k_cross}
\end{figure}

Finally, let us turn to the cross sections. One of the crucial channels for
$\bar{K}$ 
production in low energy heavy ion collisions is the strangeness exchange
reaction
\be
Y + \pi \leftrightarrow  \bar{K}+N 
\label{eq:s_exch}
\ee
For example, the $\Lambda(1405)$ resonance couples to both channels, and as
this resonance moves in mass as a result of Pauli blocking, this cross section
may change considerably \cite{schaffner,ohnishi}. Actually, as shown in 
\cite{ohnishi}, this may explain the observed pion distribution from stopped
anti-kaon reactions. In Fig.\ref{fig:k_cross} we show the results of two
different calculations for the in-medium change of the cross section for the 
reaction \raf{eq:s_exch}. In both cases, the cross sections are enhanced, 
even though the calculations and results differ in the details.

While the general trends of all calculations concerning the the in-medium 
properties of anti-kaons agree qualitatively, there are obvious model
dependencies. Those, be it the momentum
dependence of the optical potential or the density/momentum dependence of the
cross sections, can only be resolved with more detailed experimental knowledge
about the underlying $\bar{K}-N$ scattering amplitudes. At present the
available data are rather poor and only better measurements of the elementary
reactions will allow  to make definitive conclusions about the rather
interesting in-medium properties exhibited in the above figures. 

\subsubsection{Experimental results and their interpretation}

\begin{figure}[htb]
\centerline{
\epsfxsize=0.5\textwidth
\epsfbox{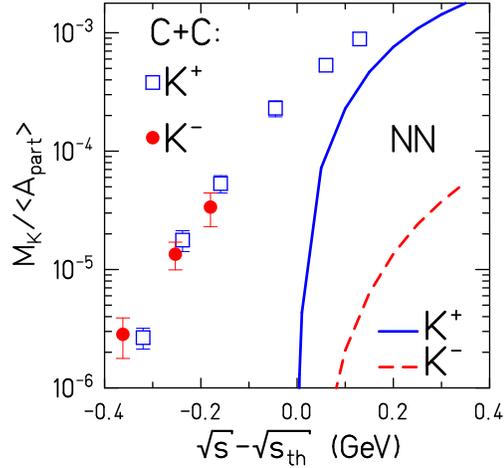}
}
\caption{Kaon and anti-kaon production as a function of energy as measured by
  the KAOS collaboation\cite{kaos}. Also shown are the $N-N$ thresholds.}
\label{fig:kaos_kbar}
\end{figure}

How are the above theoretical considerations born out in experimental data? As
already discussed in the previous section, kaons appear to fit into a rather
consistent framework. As for anti-kaons, the situation is, as expected,
considerably more complex. Most prominent is the discovery by the KAOS
collaboration \cite{kaos}, that the production of anti-kaons in heavy ion collisions
compared to proton-proton collisions is much more enhanced then that of
kaons. Indeed if plotted with respect to the threshold energy, the excitation
function of the anti-kaons and kaons are identical (see
Fig.\ref{fig:kaos_kbar}). Initially this observation gave rise to speculations
about reduction of in-medium kaon masses, 
as predicted by simple first order chiral
perturbation theory. However, from the above considerations, this scenario
seems too naive. In particular, the kaons observed by the KAOS collaboration
have a momentum larger than approx. $300 \, \rm MeV$ 
in the matter rest frame. At these
momenta, all microscopic calculations predict reduced if not vanishing 
attraction. 
 
In addition, microscopic models predict a considerable modification of the
cross section for the most important productions channel \raf{eq:s_exch}.
So what is the correct interpretation of these data? Most disturbingly,
at present transport models, 
which in this energy regime often serve as reference, do not
agree on the anti-kaons yield even absent any in-medium modifications. While
the calculations of \cite{schaffner,oeschler} more or less reproduce the
measured $K^-$ yield, the calculations of \cite{cassing,ko} definitely call
for some additional source of $K^-$. Fortunately, these discrepancies are
presently being investigated \cite{aichelin_private}.

But there is additional experimental information, which may guide us
independent of the details of the transport model results. Namely, the kaon to
anti-kaon ratio is essentially independent of centrality. How can this be
understood in terms of density dependent in-medium effects. The densities
reached in central collisions are considerably higher then those in more
peripheral ones. Maybe there is a much simpler explanation to the anti-kaons
production, as proposed by Oeschler \cite{oeschler_sqm}.

Suppose the strangeness exchange reaction as given by Eq.\ref{eq:s_exch} is
fast compared to the lifetime of the system. Then a {\em relative} chemical
equilibrium between hyperons and anti-kaons is established, with $N(Y) \gg
N(\bar{K})$.  Furthermore, in the
initial hard collisions strangeness conservation requires that as many hyperons
as kaons are produced. Consequently, given the number of kaons, the number of
hyperons is determined and in turn the number of anti-kaons. Note that this
process does not require any in-medium modifications and is consistent with
the observation \cite{redlich} that the anti-kaon yields follow a thermal
model. 

\begin{figure}[htb]
\centerline{
\epsfxsize=0.6\textwidth
\epsfbox{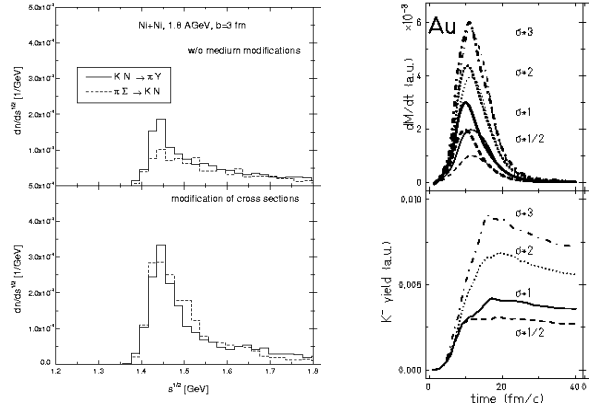}
}
\caption{Left: Number of collisions $Y+\pi \leftrightarrow K^- N$ 
\cite{schaffner} as a function of the c.m. energy. Right: Number of collisions
$Y+\pi \leftrightarrow K^- N$ \cite{oeschler} as a function of time (top) and
number of $K^-$ for different cross sections (bottom).}
\label{fig:rates}
\end{figure}

Indeed transport calculations give support to this picture. As shown in
Fig.\ref{fig:rates}, the rates for both directions of the reaction 
\raf{eq:s_exch} are almost identical, meaning {\em relative} chemical
equilibrium between hyperons and anti-kaons is essentially established. As an
interesting consequence an increase of the cross section for these
transition has only limited effect on the final anti-kaon yield
\cite{schaffner,oeschler}. As shown in the lower part of the right panel of
Fig.\ref{fig:rates} increasing the cross section by a factor of three does not
enhance the $K^-$ yield by the same amount.

\begin{figure}[htb]
\epsfxsize=0.6\textwidth
\centerline{\epsfbox{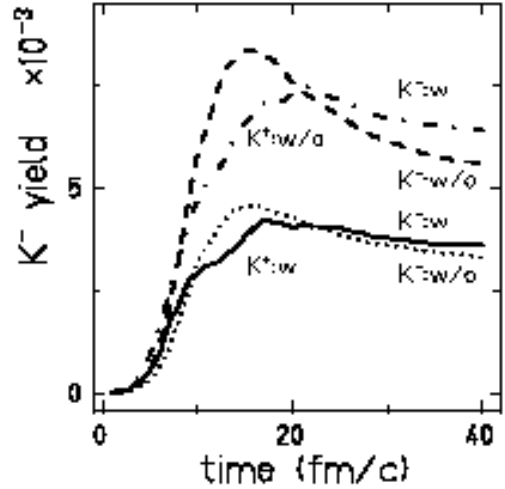}}
\caption{Number of $K^-$ for different choices of kaon and anti-kaon mean
  field potentials \cite{oeschler}.}
\label{fig:amuse}
\end{figure}

There 
is actually an even more amusing consequence of this scenario. As shown in
Fig.\ref{fig:amuse}, in-medium potentials for {\em kaons} have a larger impact
on the {\em anti-kaon} yield than in-medium potentials for the anti-kaons
themselves. This is again a simply corollary of the relative chemical
equilibrium between hyperons and kaons. Given the number of hyperons, the
anti-kaon 
number is fixed at a rather late stage of the collision. The number of
hyperons, however, is directly linked to the number of kaons and is
established early on in the dense phase of the collision. Thus, a repulsive
kaon potential, which reduces the kaon yield, affects the final anti-kaon
yield much more effectively than an anti-kaon potential.  

Another, more sobering consequence of this scenario is that heavy ion reaction
may actually not be the right environment to study the rather rich in-medium
properties of anti-kaons. If equilibrium physics 
dominates, then the subtle dynamics are hard to unravel. 
Anti-kaon nucleus scattering experiments,
therefore, seem to be 
the better environment to study this interesting question. 

In summary, the picture for the in-medium behavior of kaons seems to be
reasonably well under control and consistent with simpler systems, such as
kaon-nucleus scattering. Anti-kaons, however, are much more complex
because of their strong mixing with baryon hole states. While the different
theoretical models agree qualitatively, 
a quantitative agreement requires better
knowledge of the underlying amplitudes. Therefore, in order to make progress
in this interesting field, detailed $K^-$-nucleon scattering data are called
for. 

\section{Other aspects: equilibrium}
As already mentioned in the introduction, it is somewhat surprising that 
the strange particles, like all other particles, seem to follow the same
universal chemical freeze-out curve which connects density dominated systems
created at SIS with temperature dominated systems at SPS and RHIC. As we
have argued before, in a dense system the properties of kaons and anti-kaons
are quite different. And indeed, taking the in-medium spectral functions
seriously, a fit to a thermal model should take these spectral distributions
into account. Simply using free particle properties does not seem to be the
correct procedure. In ref. \cite{tolos_2}  an attempt in this direction has
been made. By weighting the in-medium spectral distribution with a thermal
distribution the authors are forced to use much lower temperatures
for the anti-kaons in order to reproduce that data. Or, in other words, they
do not find chemical equilibrium in these reactions. However, one
should interpret these result with some caution. The in-medium spectral
function for the anti-kaons contains considerable strength form hyperon-hole
pairs. What happens to this strength as the system expands? This will depend
on the time scales involved. In the adiabatic limit, the spectral distribution
de-mixes and all the strength will accumulate  in the kaon branch. 
However, if the expansion is sudden, some of the strength will remain in the
hyperon-hole branch, which then materializes as hyperons in the final
state. Therefore, a more detailed dynamical calculation is required before any
definite conclusions can be drawn about the effect 
of in-medium spectral functions
on the final state anti-kaons number.    

Another surprising aspect of the success of the thermal model is the fact that
the kaon should be in chemical equilibrium with the rest of the matter. Kaons
interact with small cross sections and it seems that the lifetime of the system
is too short for any equilibrium to establish \cite{mosel}. 
Luckily, kaons at SIS energies 
represent one of the few systems, where the degree of equilibrium can be
determined experimentally. As discussed in \cite{jeon} the number of kaon
pairs, though difficult to determine experimentally, provides a direct
measurement on the degree of equilibration. 
This is essentially due to the fact, that at
SIS energies strangeness is rare and thus needs to be conserved explicitly and
not only on the average as it is usually done in the grand-canonical
approximation \cite{lin}. At higher energies similar arguments can be made,
but much more complicated measurements are needed, since  strangeness is
not rare anymore \cite{abhijit}. If indeed, the measurement of the kaon pairs
establishes chemical equilibrium, then additional processes, such as many body
collisions or possible new in-medium effects need to be invoked 
in order to explain this. Thus, potentially new physics may emerge from these investigations.

\section{Summary}
In this contribution we have discussed strangeness at SIS energy heavy ion
collisions. We have concentrated on the in-medium properties of kaons and
anti-kaons and have pointed out the conceptual difference between the two in a
density dominated environment. 

In particular the anti-kaons in matter are an interesting example
for in-medium modifications via coupling to hyperon-hole modes etc. Since
strangeness is involved, the number of possible channels is limited and thus a
quantitative understanding of these effects in the future is
conceivable. However, in order to make progress, detailed information about
the basic $\bar{K}-N$ system is needed. We have also pointed out that heavy
ion collisions may not be the most appropriate tool to study this system,
since efficient partial equilibrium may wash out most of the dynamics. It
rather appears that $K^-$-nucleus scattering experiments are more appropriate
to investigate all the interesting phenomena associated with the in-medium
propagation of a hadron.    

Finally, we have mentioned that kaons at SIS energies may actually be one of
the few examples where the question of equilibrium can be addressed
in experiment. 
In case of the anti-kaons, the subtle coupled channel dynamics
needs to be controlled before this issue can be addressed with confidence.

\ack
I would like to thank M. Lutz and C. Sturm for providing some of the 
figures and J. Aichelin for many useful discussions on the status of the
transport calculations. This work was supported by the Director, 
Office of Science, Office of High Energy and Nuclear Physics, 
Division of Nuclear Physics, and by the Office of Basic Energy
Sciences, Division of Nuclear Sciences, of the U.S. Department of Energy 
under Contract No. DE-AC03-76SF00098.

\end{document}